\documentclass[prb,showpacs,floatfix,twocolumn,superscriptaddress,byrevtex]{revtex4}
\usepackage{dcolumn}
\usepackage{graphicx}
\usepackage{float}
\begin{document}

\bibliographystyle{apsrev}

\preprint{Draft version, not for distribution}
%
% The title and the list of authors
%
\title{Optical determination of the superconducting energy gap in
electron-doped \boldmath Pr$_{1.85}$Ce$_{0.15}$CuO$_4$ \unboldmath}

\author{C. C. Homes}
\email{homes@bnl.gov}%
\affiliation{Condensed Matter Physics \& Materials Science Department,
Brookhaven National Laboratory, Upton, New York 11973, USA}%
\author{R. P. S. M. Lobo}
\affiliation{Laboratoire de Physique du Solide (UPR 5 CNRS) ESPCI,
10 Rue Vauquelin 75231 Paris, France}%
\author{P. Fournier}%
\affiliation{D\'{e}partement de Physique, Universit\'{e} de Sherbrooke,
Sherbrooke, Qu\'{e}bec, J1K 2R1 Canada.}
\author{A. Zimmers}
\author{R. L. Greene}
\affiliation{Center for Superconductivity Research, Department of Physics,
University of Maryland, College Park, MD 20742, USA}%

\date{\today}

%
% The abstract goes here
%
\begin{abstract}
The  optical properties of single crystal Pr$_{1.85}$Ce$_{0.15}$CuO$_4$ have
been measured over a wide frequency range above and below the critical
temperature ($T_c \simeq 20$~K).  In the normal state the coherent part of the
conductivity is described by the Drude model, from which the scattering rate
just above $T_c$ is determined to be $1/\tau \simeq 80$~cm$^{-1}$.  The
condition that $\hbar/\tau \approx 2k_B T$ near $T_c$ appears to be a general
result in many of the cuprate superconductors.  Below $T_c$ the formation of a
superconducting energy gap is clearly visible in the reflectance, from which
the gap maximum is estimated to be $\Delta_0 \simeq 35$~cm$^{-1}$ (4.3~meV).
The ability to observe the superconducting energy gap in the optical properties
favors the nonmonotonic over the monotonic description of the {\it d}-wave gap.
The penetration depth for $T\ll T_c$ is $\lambda \simeq 2000$~\AA , which when
taken with the estimated value for the dc conductivity just above $T_c$ of
$\sigma_{dc} \simeq 35 \times 10^3$ $\Omega^{-1}$cm$^{-1}$ places this material
on the general scaling line for the cuprates defined by $1/\lambda^2 \propto
\sigma_{dc}(T\simeq T_c) \cdot T_c$.  These results are consistent with the
observation that $1/\tau \approx 2\Delta_0$, which implies that the material is
not in the clean limit.
\end{abstract}
%
%  PACS numbers
%  63.20.-e  Phonons in crystal lattices
%  72.15.Lh  Relaxation times and mean-free paths
%  74.25.Gz  Optical properties
%  74.25.-q  Normal and superconducting states
%  74.72.-h  High-Tc compounds
%  74.72.Bk  Y-based cuprates
%  77.22.Ch  Permittivity (dielectric function)
%  78.30.-j  Infrared and Raman spectra
%
\maketitle

%
% The main body of the text
%
% Introduction
%
\section{Introduction}
The high-temperature copper-oxide superconductors may be grouped into two broad
categories; hole- and electron-doped materials.  The hole-doped materials
constitute the majority of the cuprate superconductors, while electron doping
is observed in a relatively small number of materials, mainly the
(Nd,Pr)$_{2-x}$Ce$_x$CuO$_4$ systems\cite{tokura89,takagi89a} and the
infinite-layer (Sr,$L$)CuO$_2$ ($L=$ La, Sm, Nd, Gd)
materials.\cite{siegrist88,smith91} The phase diagrams of the hole- and
electron-doped materials have some similarities,\cite{damascelli03} with the
parent materials being antiferromagnetic (AFM) insulators in both cases.
However, the electron-doped materials are also noticeably different in that the
AFM region extends to a much higher doping with an almost non-existent
pseudogap region.\cite{luke90,timusk99,mang04} In addition, the superconducting
dome is quite small, with relatively low critical temperatures ($T_c$'s). These
differences have prompted some debate as to whether or not the electron-doped
materials were high-temperature superconductors at all, or if they resembled
more conventional superconductors. While some work indicated that the
electron-doped materials possess an isotropic {\it s}-wave superconducting
energy gap,\cite{huang90,stablober95} more recent studies have suggested that
the energy gap has nodes and is {\it d}-wave in
nature,\cite{tsuei00b,biswas01,armitage01,sato01,kokales00,prozorov00a,skinta02,
blumberg02,blumberg03,matsui05b,shan05} similar to the hole-doped
materials.\cite{harlingen95,tsuei00a}  However, while the {\it d}-wave gap in
the hole-doped materials may be described in a monotonic way, $\Delta(\phi) =
\Delta_0 \cos(2\phi)$, where $\Delta_0$ is the gap maximum and $\phi$ is a
Fermi surface angle, the energy gap in the electron-doped materials appears to
be nonmonotonic, i.e., $\Delta(\phi) = \Delta_0 \left[ \cos(2\phi) -
a\cos(6\phi) + b\cos(10\phi) \right]$.\cite{blumberg02,blumberg03,matsui05b}

In electron-doped materials, a pseudogap exists in the underdoped regime and
overlaps superconductivity over a small portion of the phase
diagram.\cite{onose01,zimmers05}  In the normal state above the superconducting
dome, the conductivity is reasonably metallic.\cite{onose01} The formation of a
superconducting energy gap and the commensurate change in the density of states
(DOS)  should lead to observable changes in the low-energy optical properties.
In initial optical studies of the electron-doped (Nd,Pr)$_{2-x}$Ce$_x$CuO$_4$
materials,\cite{cooper90,zhang91,lupi92,arima93,homes97,onose99,singley01,onose01,
onose04,wang06} there was no definitive signature in the reflectance of a gap
opening.  It was suggested that this was due to the fact that the
high-temperature superconductors were in the clean limit (i.e., the regime
where the normal-state scattering rate $1/\tau \ll \Delta_0$, or alternatively
when the mean free path is much greater than the coherence length, $l \gg
\xi_0$), where the formation of a superconducting gap is difficult to
observe.\cite{kamaras90} Interestingly, we have recently observed changes in
the reflectance of thin films of Pr$_{2-x}$Ce$_x$CuO$_4$ above and below $T_c$
which track with doping,\cite{zimmers04} indicating that these features are
associated with the superconducting energy gap.  However, in any study of thin
films there is always the concern that substrate-induced strain may affect the
structural and electronic properties of the film.

%
% Results of this study
%
In this work we examine the optical properties of an optimally-doped single
crystal of Pr$_{1.85}$Ce$_{0.15}$CuO$_4$ ($T_c \simeq 20$~K) for light
polarized in the copper-oxygen planes over a wide frequency range in the normal
and superconducting states.  Some aspects of this work have been previously
reported.\cite{millis05}  The results show a clear signature of the formation
of a superconducting energy gap in the reflectance and the optical
conductivity, validating the earlier thin-film work.\cite{zimmers04} The
normal-state properties are well described by a simple two-component model
(coherent and incoherent components), which allows the plasma frequency of the
coherent Drude component $\omega_{pd}$ and the scattering rate $1/\tau$ to be
determined.
At 30~K, $1/\tau \simeq 80$~cm$^{-1}$ or about 10~meV, which is consistent with
the observation that $\hbar/\tau \approx 2k_B T$ near $T_c$ in many of the
cuprate superconductors.   In the superconducting state, the real part of the
dielectric function and optical conductivity sum rules both yield estimates for
the in-plane penetration depth of $\lambda \simeq 2000$~\AA .
From the structure observed in the reflectance below $T_c$, the gap maximum is
estimated to be $\Delta_0 \simeq 35$~cm$^{-1}$ (about 4.3~meV).  For values of
$1/\tau$ determined just above $T_c$, the result that $1/\tau \approx
2\Delta_0$ (valid primarily along the antinodal direction) indicates that the
material is not in the clean limit, in agreement with recent scaling
arguments.\cite{homes04,homes05}  In addition, we speculate that the
nonmonotonic nature of the superconducting gap in this material results in
large changes to the joint density-of-states (JDOS) below $T_c$ relative to the
monotonic case, allowing the formation of the gap to be observed more easily in
the optical response.

%
% Experimental section
%
\section{Experimental}
Single crystals of Pr$_{1.85}$Ce$_{0.15}$CuO$_4$ were grown from a CuO-based
flux using a directional solidification technique.\cite{peng91}  A mixture of
high-purity (99.9\%) starting materials of Pr$_6$O$_{11}$, CeO$_2$ and CuO were
heated rapidly to just above the melting point ($\sim 1270^\circ$C for this Ce
concentration). After a soak of several hours at the maximum temperature, the
materials were cooled slowly to room temperature. To induce superconductivity,
the crystals (typical size of $2\times 2$~mm$^2$ and $\sim 20\,\mu$m thickness)
were oxygen reduced by annealing in an inert gas atmosphere following a
procedure similar to that described by Brinkmann {\it et al.}\cite{brinkmann96}
The superconducting transition was characterized by a SQUID magnetometer from
Quantum Design in a field of 1~Oe (ZFC), and the critical temperature
determined to be $T_c \simeq 20$~K. The observed value of $T_c$ for this Ce
concentration is somewhat less than ideal, suggesting that the sample may have
been over-reduced.

The reflectance of single-crystal Pr$_{1.85}$Ce$_{0.15}$CuO$_4$ has been
measured at a near-normal angle of incidence for light polarized in the {\it
a-b} planes from $\approx 18$ to over $34\,000$~cm$^{-1}$, above and below
$T_c$, on Bruker IFS 66v/S and IFS 113v spectrometers using an {\it in-situ}
evaporation technique.\cite{homes93} The noise in the far-infrared reflectance
is less than 0.05\%, resulting in a signal-to-noise ratio of better than
$2000\!:\!1$. The optical properties are calculated from a Kramers-Kronig
analysis of the reflectance, where extrapolations are supplied for $\omega
\rightarrow 0, \infty$.  At low frequency, a metallic Hagen-Rubens response is
assumed in the normal state $(R \propto 1-\omega^{1/2})$, and a two-fluid model
was applied in the superconducting state $(R \propto 1-\omega^2)$.  Above the
highest-measured frequency in this experiment the reflectance of
Pr$_{1.85}$Ce$_{0.15}$CuO$_4$ has been employed to about 35~eV;\cite{arima93}
above that frequency a free-electron approximation has been assumed $(R\propto
1/\omega^4)$.

%
% Figure 1
%
\begin{figure}[t]%
%
% eprint, manuscript
%
\vspace*{-0.5cm}%
\centerline{\includegraphics[width=3.6in]{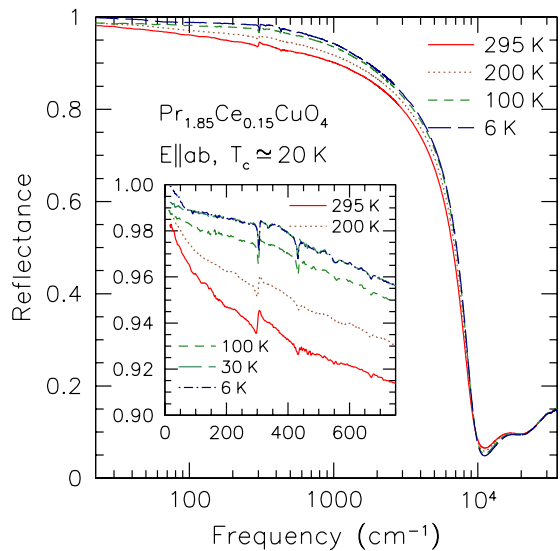}}%
%\centerline{\includegraphics[width=4.6in]{figure1.eps}}%
\vspace*{-1.4cm}%
\caption{(Color online) The temperature dependence of the reflectance at a
near-normal angle of incidence of single crystal Pr$_{1.85}$Ce$_{0.15}$CuO$_4$
($T_c \simeq 20$~K) from $\approx 18$ to $34\,000$~cm$^{-1}$  for light
polarized in the {\it a-b} plane above and below $T_c$.
Inset: The detailed temperature dependence of the far-infrared reflectance
above and below $T_c$.  Note the kink and the sudden increase in the
reflectance below $\simeq 70$~cm$^{-1}$ for $T \ll T_c$.  The sharp features in
the reflectance are infrared-active lattice modes. (The resolution is
2~cm$^{-1}$.)}
\vspace*{-0.3cm}%
\label{fig:reflec}
\end{figure}

%
% Results and discussion
%
\section{Results and Discussion}
\subsection{Optical and transport properties}
The {\it ab}-plane reflectance of Pr$_{1.85}$Ce$_{0.15}$CuO$_4$ ($T_c \simeq
20$~K) is shown in Fig.~\ref{fig:reflec} over a wide spectral range, at a
variety of temperatures above and below $T_c$. The reflectance in the
mid-infrared region and above is consistent with previous studies of
electron-doped cuprates;\cite{cooper90,zhang91,lupi92,arima93,homes97,
onose99,singley01,onose01,onose04,wang06} there is a plasma edge at $\approx
1.2$~eV, and the reflectance throughout the mid-infrared region increases with
decreasing temperature.  The inset in Fig.~\ref{fig:reflec} shows the
low-frequency reflectance at several temperatures above and below $T_c$.  The
sharp structures in the reflectance that appear variously as resonant and
antiresonant features are infrared-active lattice modes.\cite{braden05} The
reflectance increases with decreasing temperature, but there is little change
between the 30 and 6~K spectra, with the exception of a kink at $\simeq
70$~cm$^{-1}$, below which the reflectance increases rapidly with decreasing
frequency. This is the same feature in the reflectance that was observed in
thin-film studies,\cite{zimmers04} that signals the formation of a
superconducting energy gap.

%
% Figure 2
%
\begin{figure}[t]%
%
% eprint, manuscript
%
\vspace*{-0.3cm}%
\centerline{\includegraphics[width=3.8in]{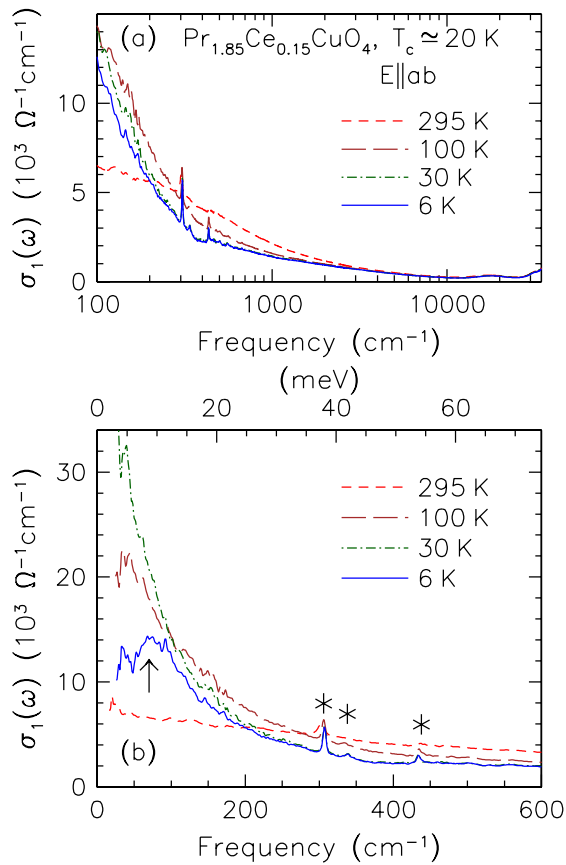}}%
%\centerline{\includegraphics[width=4.6in]{figure2.eps}}%
\vspace*{-1.0cm}%
\caption{(Color online) (a) The temperature dependence of the real part of the
optical conductivity over a wide frequency range for
Pr$_{1.85}$Ce$_{0.15}$CuO$_4$ above and below $T_c$ for light polarized in the
{\it a-b} planes.
(b) The low-frequency optical conductivity. The conductivity in the normal
state is described by a Drude component that narrows rapidly with decreasing
temperature.  Below $T_c$, there is a significant loss of spectral weight at
low frequency (arrow) due to the formation of a condensate. The sharp features
superimposed upon the electronic background at $\simeq 306$, 338, and
433~cm$^{-1}$ (asterisks) are infrared-active phonon modes.\cite{braden05} }
\vspace*{-0.3cm}%
\label{fig:sigma}
\end{figure}

The optical conductivity is shown over a wide frequency range in
Fig.~\ref{fig:sigma}(a), and at low frequency in Fig.~\ref{fig:sigma}(b). The
conductivity in the normal-state can be described as a combination of a
coherent Drude component that describes the far infrared response, and an
incoherent component that dominates the mid infrared region. The
``two-component'' expression for the real part of the optical conductivity is
\begin{equation}
  \sigma_1(\omega) = {1\over{4\,\pi}} \, { {\omega_{pd}^2 \Gamma} \over
   {\omega^2+\Gamma^2} }+\sigma_{\rm MIR},
   \label{eq:sigma}
\end{equation}
where $\omega_{pd}^2 = 4\pi n_d e^2/m^\ast$ is the square of the Drude plasma
frequency, $n_d$ is a carrier concentration associated with coherent transport,
$m^\ast$ is an effective mass, $\Gamma=1/\tau$ is the scattering rate, and
$\sigma_{\rm MIR}$ is the mid-infrared component.  [When $\omega_{pd}$ and
$1/\tau$ are in cm$^{-1}$, $\sigma_1$ also has the units cm$^{-1}$; to recover
units for the conductivity of $\Omega^{-1}$cm$^{-1}$, the term $1/4\pi$ in
Eq.~(\ref{eq:sigma}) should be replaced with $2\pi/Z_0$, where
$Z_0=377\,\Omega$ is the characteristic impedance of free space.] The Drude
contribution has the form of a Lorentz oscillator centered at zero frequency.
The features in the reflectance attributed to lattice modes appear as sharp
resonances in the conductivity, shown in detail in Fig.~\ref{fig:sigma}(b).  To
apply the two-component model to the data, it is necessary to specify the
nature of $\sigma_{\rm MIR}$. The mid-infrared conductivity is often described
by a series of overdamped Lorentzian oscillators which yield a flat, incoherent
response in this region. However, this approach can be somewhat arbitrary.  To
simplify the fitting a constant background ($\sigma_{\rm MIR}\simeq 1000$
$\Omega^{-1}{\rm cm}^{-1}$) has been used at low temperature.
The fitted results show that while the scattering rate decreases from $1/\tau =
130\pm 3$ to $80\pm 2$ ~cm$^{-1}$ when the temperature is reduced from 100 K to
just above $T_c$ at 30~K, the Drude plasma frequency remains constant at
$\omega_{pd} = 13\,000\pm 200$~cm$^{-1}$.   Transport measurements in the
electron-doped cuprates typically show a quadratic form of the
resistivity,\cite{jiang94,peng97,fournier97,skinta02,dagan04} $\rho = \rho_0 +
aT^2$, with a weak temperature dependence near $T_c$, indicating that at low
temperatures $1/\tau$ is dominated by $\rho_0$. Interestingly, the result that
$\hbar/\tau \approx 2k_B T$ close to $T_c$ seems to be true for many of the
cuprates.

%
% Total number of doped carriers relative to the coherent component
%
Note that $\omega_{pd}^2$ is a reflection of only those carriers that
participate in coherent transport, rather than the total number of doped
carriers determined from the classical plasma frequency $\omega_p^2=4\pi n
e^2/m$. The value for $\omega_p$ may be estimated using several different
techniques. The zero-crossing of the real part of the dielectric function
$\epsilon_1(\omega)$ occurs at $\omega_p/\sqrt{\epsilon_\infty}$.  However, the
presence of interband absorptions that overlap with the coherent component, as
well as the difficulty in choosing the correct value of $\epsilon_\infty$,
makes this approach unreliable.
Another method is the finite-energy $f$-sum rule\cite{smith}
\begin{equation}
  \int_0^{\omega_c} \sigma_1(\omega) d\omega \approx \omega_p^2/8,
\end{equation}
where $\omega_c$ is a cut-off frequency.  In the absence of other excitations,
this sum rule is exact in the $\omega_c \rightarrow \infty$ limit.  While this
condition is difficult to achieve in the cuprates, modifications to this sum
rule based on an analysis of the absorption coefficient $\alpha(\omega)$ have
been suggested by Hwang {\it et al.}\cite{hwang04} Adopting this approach
yields $\omega_p \simeq 19\,300$~cm$^{-1}$, suggesting that only about half of
the doped carriers participate in coherent transport (assuming the masses do
not change), similar to the hole-doped materials.\cite{tanner98}

\begin{figure}[t]%
%
% eprint, manuscript
%
\vspace*{-0.5cm}%
\centerline{\includegraphics[width=3.6in]{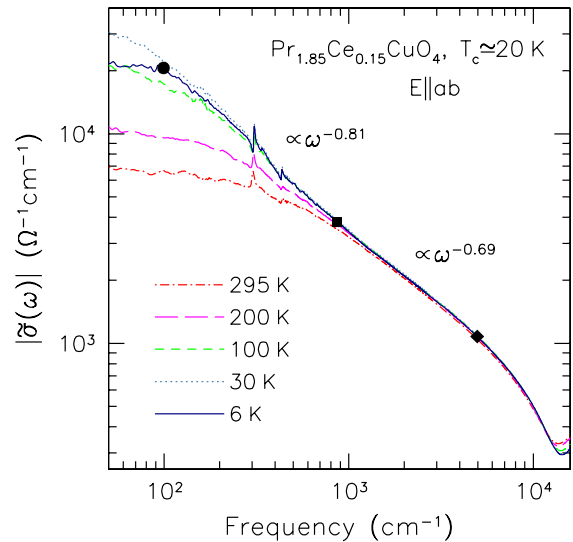}}%
%\centerline{\includegraphics[width=4.6in]{figure3.eps}}%
\vspace*{-1.4cm}%
\caption{(Color online) The log-log plot of the temperature dependence of the
modulus of the optical conductivity over a wide frequency range above and below
$T_c$.  Throughout the mid-infrared region (between the square and the
diamond), the modulus obeys a power-law behavior, $|\tilde\sigma(\omega)|
\propto \omega^{-0.7}$.  At low temperature the power-law behavior is recovered
again at low frequency (between the square and the dot),
$|\tilde\sigma(\omega)| \propto \omega^{-0.8}$.  From the relation
$\tilde\sigma(\omega) = \sigma_1(\omega) + i\sigma_2(\omega) = -i\,\omega [
\tilde\epsilon(\omega) - \epsilon_\infty ]/4\pi$, a value of $\epsilon_\infty =
3.6$ was used to determine $\sigma_2(\omega)$.] }
\vspace*{-0.3cm}%
\label{fig:mod}
\end{figure}

%
% Talk about the power-law behavior
%
While the majority of this paper is concerned with the far-infrared optical
properties, it is worth commenting briefly on the behavior of the optical
conductivity throughout the rest of the infrared frequency range.  As
Fig.~\ref{fig:sigma}(a) indicates, at room temperature the conductivity is
quite broad, with little frequency dependence at low energies; however, as the
temperature decreases the Drude component narrows rapidly.  The reduction in
$1/\tau$ leads to changes in the conductivity over much of the far infrared;
however, with the exception of a weak feature at $\simeq 18\,000$~cm$^{-1}$,
the high-frequency conductivity displays little structure.  While this result
is consistent with an earlier study of single crystal
Nd$_{1.85}$Ce$_{0.15}$CuO$_4$ grown by a flux technique,\cite{homes97} other
works on crystals grown using the traveling-solvent floating-zone method have
revealed some unusual structure in the $300 - 400$~cm$^{-1}$ ($40 - 50$~meV)
region.\cite{singley01,onose01, onose04}  The as-grown electron-doped materials
are not superconducting; they must be oxygen-reduced to induce a $T_c$.
Interestingly, the as-grown (or oxygenated) samples show prominent structure in
the reflectance in the $300 - 400$~cm$^{-1}$ region that manifests itself as a
suppression of the conductivity, that is almost completely removed upon oxygen
reduction.\cite{onose99,singley01}  We speculate that the differences observed
in the various works may be related to different levels of oxygen reduction.
As previously mentioned, the conductivity is reasonably well described in the
far-infrared region by a Drude response. However, it has been noted in the
hole-doped cuprates that the modulus of the conductivity obeys a power law over
much of the mid-infrared region,\cite{azrak94,marel03} $|\tilde\sigma(\omega)|
\propto \omega^{-0.65}$.  The log-log plot of the temperature dependence of the
modulus of the optical conductivity vs frequency is shown in Fig.~\ref{fig:mod}
for a variety of temperatures over a wide frequency range. Throughout much of
the mid infrared, the modulus of the conductivity follows the power law
$|\tilde\sigma(\omega)| \propto \omega^{-0.69}$, in good agreement with the
behavior observed in the hole-doped
cuprates.\cite{azrak94,anderson97,marel03,norman06,krotkov06a,krotkov06b} Below
roughly 1000~cm$^{-1}$ there is a deviation from this power-law behavior at
high temperature.  As the temperature decreases a linear behavior is once again
recovered; however, the exponent is now larger $|\tilde\sigma(\omega)| \propto
\omega^{-0.81}$, suggesting that the character of the conductivity is different
in these two regions.
%
% Figure 3
%

\begin{figure}[t]%
%
% eprint, manuscript
%
\vspace*{-0.5cm}%
\centerline{\includegraphics[width=3.6in]{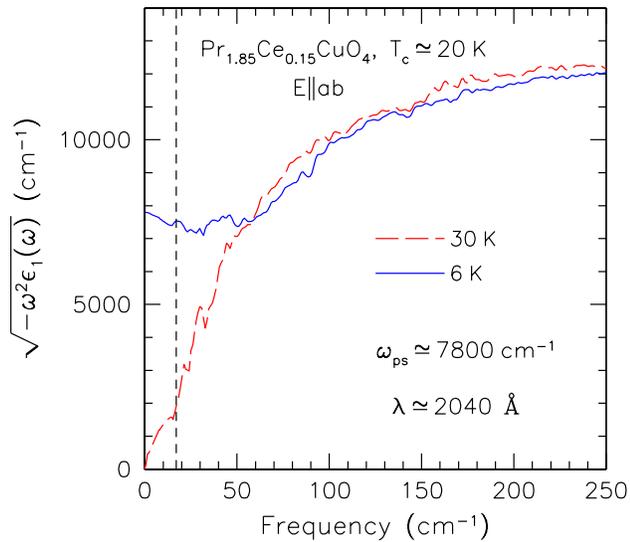}}%
%\centerline{\includegraphics[width=4.6in]{figure4.eps}}%
\vspace*{-1.4cm}%
\caption{(Color online) The temperature dependence of $\sqrt{-\omega^2
\epsilon_1(\omega)}$ of Pr$_{1.85}$Ce$_{0.15}$CuO$_4$ at 30 and 6~K. The
results of the low-frequency extrapolations employed in the Kramers-Kronig
analysis below $\simeq 18\,$cm$^{-1}$ (dashed line) are included as a guide to
the eye. }
\vspace*{-0.3cm}%
\label{fig:rho}
\end{figure}

Well below $T_c$, there is a substantial reduction in the low-frequency
conductivity [indicated by the arrow in Fig.~\ref{fig:sigma}(b)].   The
Ferrell-Glover-Tinkham (FGT) sum rule\cite{ferrell58, tinkham59} states that
the difference between the conductivity curves at $T \simeq T_c$ and $T\ll T_c$
(the so-called ``missing area'') is related to the formation of a
superconducting condensate
\begin{equation}
  \int_{0^+}^{\omega_c} \left[ \sigma_{1}(\omega,T\simeq T_c) -
  \sigma_{1}(\omega,T\ll T_c) \right] \approx \omega_{ps}^2/8,
\end{equation}
where $\omega_{ps}^2=4\pi n_s e^2/m^\ast$ is the square of the superconducting
plasma frequency, $n_s$ is the superconducting carrier concentration, and
$\omega_c$ is chosen such that $\omega_{ps}^2$ converges smoothly.  The
strength of the condensate is simply $\rho_s = \omega_{ps}^2$, which is related
to the penetration depth by $\rho_s = c^2/\lambda^2$.
The value of $\rho_s$ may also be estimated from the response of the dielectric
function in the zero-frequency limit to the formation of a condensate, which is
expressed purely by the real part $\epsilon_1(\omega\rightarrow 0) =
\epsilon_\infty -\omega_{ps}^2 / \omega^2$ for $T\ll T_c$.  This allows the
strength of the condensate to be calculated from $\omega_{ps}^2 \simeq
-\omega^2 \epsilon_1(\omega)$ as $\omega \rightarrow 0$. The frequency
dependence of $\sqrt{-\omega^2 \epsilon_1(\omega)}$ is shown in
Fig.~\ref{fig:rho} for Pr$_{1.85}$Ce$_{0.15}$CuO$_4$ at 30 and 6~K. The
low-frequency extrapolations employed in the Kramers-Kronig analysis of the
reflectance are included to allow the $\omega \rightarrow 0$ values to be
determined more easily.  In the normal state the function goes smoothly to
zero, indicating the absence of a condensate.  Well below $T_c$, the estimate
for the superconducting plasma frequency is $\omega_{ps}\simeq 7800$~cm$^{-1}$.
This is consistent with values of $\omega_{ps}\simeq 8000$~cm$^{-1}$ determined
from the FGT sum rule. From $1/\lambda=2\pi\omega_{ps}$ the penetration depth
is determined to be $\lambda = 2000\pm 100$~\AA , similar to results obtained
from thin films with similar $T_c$'s.\cite{prozorov00b,zimmers04}
A comparison of $\omega_{ps}$ to $\omega_{pd}$ indicates that less than half of
the carriers involved in the coherent Drude component have collapsed into the
condensate (assuming similar effective masses), a result that is consistent
with the larger body of work on the hole-doped materials.\cite{tanner98}

%
% Figure 4
%
%

%
% Comment about scaling
%
The relatively low values for $T_c$ and $\lambda$ for this material, and the
electron-doped materials in general, have always presented a challenge for the
Uemura plot,\cite{uemura88} which relates the density of carriers in the
superfluid to the transition temperature, $\rho_s \propto T_c$. While the
Uemura plot works well for the hole-doped cuprates in the underdoped region,
the electron-doped materials have typically fallen well off of the Uemura
plot.\cite{homes97,shengelaya05,homes06} Recently, a more general scaling
relation $\rho_s \propto \sigma_{dc}\,T_c$ (where $\sigma_{dc}$ is determined
just above $T_c$) which allows the points for the electron- and hole-doped
cuprates to be scaled onto the same universal line.\cite{homes04,homes05} The
value of $\sigma_{dc}$ close to $T_c$ is taken from both
$\sigma_1(\omega\rightarrow 0)$ as well as Drude fits to the lineshape of the
conductivity, resulting in the estimate $\sigma_{dc}(T\approx T_c) = 35\,000\pm
3000$~$\Omega^{-1}$cm$^{-1}$; this places the material almost directly on the
$\rho_s \propto \sigma_{dc}\,T_c$ scaling line.  An implication of this result
is the suggestion that this material is not in the clean limit;\cite{homes05}
this controversial point will be examined in more detail in a subsequent
section.

%
% Subsection - Determination of the superconducting energy gap
%
%
\subsection{Determination of the gap maximum}
\begin{figure}[t]%
%
% eprint, manuscript
%
\vspace*{-0.7cm}%
\centerline{\includegraphics[width=3.6in]{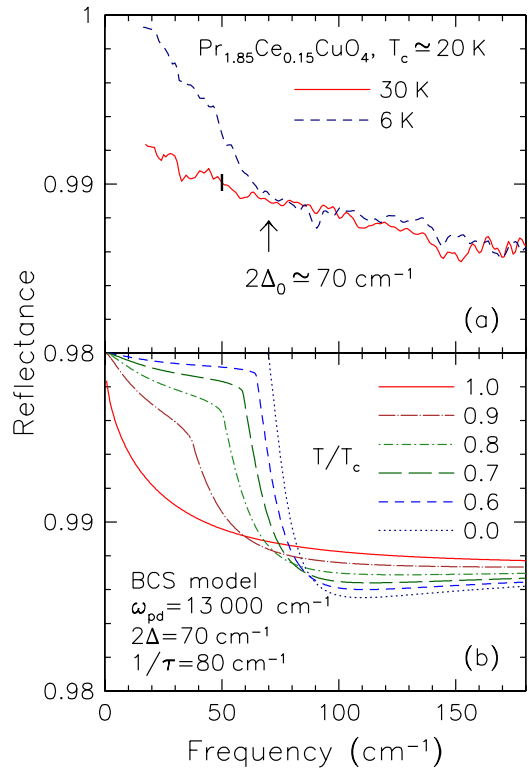}}%
%\centerline{\includegraphics[width=4.6in]{figure5.eps}}%
\vspace*{-1.2cm}%
\caption{(Color online) (a) The far-infrared reflectance of
Pr$_{1.85}$Ce$_{0.15}$CuO$_4$ above and below $T_c$.  The kink in the
reflectance below $T_c$ signals the formation of a superconducting energy gap
and denotes the value of $2\Delta_0$.  The estimated noise in the reflectance
is indicated by the thick line at 50~cm$^{-1}$; the S/N is in excess of
$2000\!:\!1$, and no smoothing has been applied to the data.
(b) The calculated reflectance of a BCS superconductor with an isotropic {\it
s}-wave gap for a series of temperatures at and below $T_c$.  }
\vspace*{-0.3cm}%
\label{fig:bcs}
\end{figure}

A reasonable estimate for the gap maximum may be taken from a comparison of the
reflectance at $T\simeq T_c$ and $T\ll T_c$, shown in Fig.~\ref{fig:bcs}(a).
Above about 70~cm$^{-1}$ the two curves are nearly identical, but below this
value the 6~K reflectance displays a kink followed by an abrupt increase; this
feature has also been observed in our measurements of the reflectance of thin
films of this material.\cite{zimmers04}  The optical conductivity (and by
extension, the reflectance) of a material may be described by the
Kubo-Greenwood formula, which considers all the single-electron transitions
across the Fermi surface, a measure of the JDOS.\cite{harrison}  It is not
possible to reproduce the structure in the reflectance simply by considering
the response of the dielectric function to the formation of a condensate;
instead, the kink is associated with a DOS effect due to the formation of a
superconducting gap. From the JDOS, the position of the kink should correspond
to twice the gap maximum. A reasonable estimate for twice the gap maximum is
therefore $2\Delta_0 \simeq 70$~cm$^{-1}$, or $\Delta_0 \simeq 4.3$~meV; this
yields a ratio of $2\Delta_0/k_BT_c \simeq 5$, in good agreement with previous
thin film\cite{zimmers04} and single crystal results.\cite{qazilbash05}
Furthermore, the value for $\Delta_0$ is in excellent agreement with the values
determined from Raman studies.\cite{blumberg02,blumberg03}
%
% Figure 5
%

\begin{figure}[t]%
%
% eprint, manuscript
%
\vspace*{-3.4cm}%
\centerline{\includegraphics[width=3.8in]{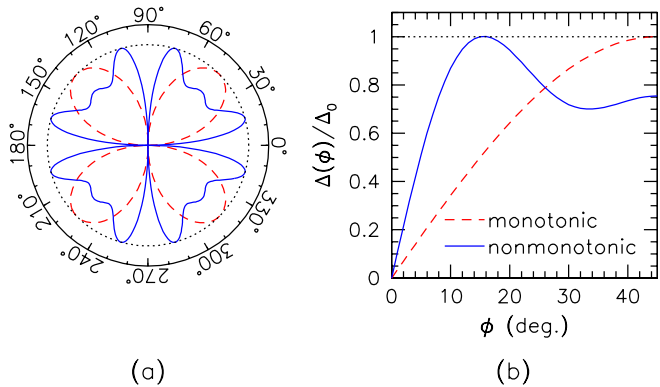}}%
%\centerline{\includegraphics[width=4.6in]{figure6.eps}}%
\vspace*{-1.0cm}%
\caption{(Color online) (a) A radial plot of the amplitude of an isotropic {\it
s}-wave gap (dotted line), a monotonic {\it d}-wave gap, $\Delta(\phi) =
\Delta_0\cos(2\phi)$ (dashed line), and a nonmonotonic {\it d}-wave gap,
$\Delta(\phi) = \Delta_0 \left[ \cos(2\phi) - 0.42\cos(6\phi) +
0.17\cos(10\phi) \right]$.  Note that the gap functions are rotated by
$45^\circ$ with respect to the hole-doped cuprates.
(b) A linear plot of the same gap functions over the first quadrant.  }
\vspace*{-0.3cm}%
\label{fig:gaps}
\end{figure}

The nature of the superconducting energy gap also plays a critical role in our
ability to observe it.  Gaps with sharp features in the DOS, and therefore the
JDOS, should have features that are more easily observed in the optical
properties.
To illustrate this point, gap functions for isotropic {\it s}-wave, monotonic
and nonmonotonic {\it d}-wave materials are shown in Fig.~\ref{fig:gaps}(a) and
$\Delta(\phi)/\Delta_0$ is shown over the first quadrant in
Fig.~\ref{fig:gaps}(b).  The nonmonotonic {\it d}-wave gap will have a gap
maximum much closer to the nodes, resulting in a larger part of the Fermi
surface that is effectively gapped.  In comparison, the monotonic {\it d}-wave
gap associated with the hole-doped cuprates has a gap maximum far from the
nodal regions and the resulting JDOS is rather smeared out.  We propose that
the non-monotonic nature of the {\it d}-wave gap in the electron-doped material
makes it possible to identify $\Delta_0$ from the optical properties.  However,
this is by no means restricted to the electron-doped materials.  There are
large changes observed in the reflectance of the hole-doped cuprates below
$T_c$, the so-called ``knee'' in the reflectance\cite{tanner92} located at
roughly $2\Delta_0$, that may be due to DOS effects related to the formation of
a superconducting energy gap.\cite{lee05} In those cases where this feature was
observed above $T_c$, it was argued that it was not related to the
superconductivity. However, many of the cuprate superconductors initially
studied were naturally underdoped; these materials display a pseudogap that
develops in the normal state.\cite{timusk99} Angle resolved photoemission
spectroscopy has demonstrated that the pseudogap entails a partial gapping of
the Fermi surface in a manner similar to that of the superconducting energy
gap.\cite{kanigel06}  In addition, the calculations of the optical conductivity
based on a monotonic {\it d}-wave gap are in excellent agreement with the
experiment.\cite{carbotte04} Thus, the appearance of the knee in the
reflectance above $T_c$ in the underdoped materials does not rule out the
association of this feature with the superconducting energy gap for $T \ll
T_c$.

%
% Simpler in electron-doped materials
%
While it is therefore possible to observe the DOS effects of the
superconducting gap in materials with a simple monotonic {\it d}-wave gap, we
argue that this task is simplified considerably if the gap is nonmonotonic. To
elaborate on this point, we calculate the temperature dependence of the
reflectance of a material using the BCS model with an isotropic {\it s}-wave
gap for an arbitrary purity level.\cite{zimmerman91}
%
%
% Figure 6
%
The normal-state is described using the Drude model with a plasma frequency of
$\omega_{pd} = 13\,000$~cm$^{-1}$ and scattering rate $1/\tau = 80$~cm$^{-1}$,
while below $T_c$ the optical properties have been calculated with a gap of
$2\Delta = 70$~cm$^{-1}$.  The calculated reflectance curves are shown in
Fig.~\ref{fig:bcs}(b).  The normal-state reflectance at $T_c$ is reproduced
quite well, while below $T_c$ the formation of an isotropic {\it s}-wave gap
produces a region of steadily increasing reflectance for $\omega \lesssim
2\Delta_0$; for $T\ll T_c$ the gap is fully formed and the reflectance is unity
below $2\Delta_0$.  While it is clear that the gap in
Pr$_{1.85}$Ce$_{0.15}$CuO$_4$ is {\it d} wave, the nonmonotonic nature of the
gap results in more of the Fermi surface being more effectively gapped than the
monotonic case (reminiscent of an isotropic gap), resulting in a JDOS which
allows for the unambiguous determination of $2\Delta_0$.

\subsection{Anisotropy and the effects of disorder}
In terms of the optically-determined values for $1/\tau$ and $2\Delta_0$, the
clean and dirty limits are defined as $1/\tau \ll 2\Delta_0$ and $1/\tau
\gtrsim 2\Delta_0$, respectively (where it is understood that ``dirty'' refers
to the effects of disorder and electronic correlations rather than impurity
effects).  Although the cuprates are generally considered to be in the clean
limit, we now are faced with the condition that $1/\tau \approx 2\Delta_0$,
which places the material close to the dirty limit.  The widely-accepted
statement that this class of materials is in the clean limit is based on the
incorrect comparison of the small value for the quasiparticle scattering along
the nodal direction and the gap value along the antinodal direction, two
different directions in momentum space.  This statement does not take into
consideration the anisotropic nature of the Fermi surface of these materials in
which both $1/\tau$ and the superconducting energy gap vary significantly,
depending on whether the nodal or antinodal directions are being considered.
%
% The scattering rate and the gap...
%
For instance, the quasiparticle scattering rate is observed to drop abruptly in
the cuprates below $T_c$ after the antinodal region of the Fermi surface is
gapped, suggesting that the scattering rate in the nodal direction is much
smaller than the antinodal direction.\cite{bonn93,valla00}  (In the underdoped
cuprates the formation of a pseudogap leads to much the same behavior in the
normal state; the gapping of the antinodal regions restricts the quasiparticles
to the nodal part of the Fermi surface, where they display a metallic
character,\cite{sutherland05,lee05} i.e., a ``nodal metal'').  In the absence
of a pseudogap, Matthiesen's rule implies that the $1/\tau$ observed in the
normal state therefore arises from scattering mainly in the antinodal
direction, and that as a consequence $1/\tau$ is anisotropic.  In addition, the
superconducting energy gap is highly-anisotropic due to its {\it d}-wave
nature, with $\Delta_0$ in this work estimated to be $\simeq 4.3$~meV.  The
comparison of the normal-state scattering rate with the superconducting energy
gap maximum correctly compares two quantities associated with the antinodal
direction.

%
% Table I - Fermi surface parameters
%
\begin{table}[tb]
\caption{The estimated values at the nodal and antinodal points of the Fermi
surface for the Fermi velocity, scattering rate and mean-free path ($T\gtrsim
T_c$); the magnitude of the superconducting energy gap and the coherence length
($T\ll T_c$).}
\begin{ruledtabular}
\begin{tabular}{cccccc}
%
%       & \multicolumn{3}{c}{$T\gtrsim T_c$} & \multicolumn{2}{c}{$T\ll T_c$} \\
       & $v_F^a$  &  $1/\tau$   & $l$   & $\Delta(\phi)$ & $\xi_{0,calc}^b$ \\
Region & ($10^7$~cm/s) & (cm$^{-1})$ & (\AA) &    (meV)       & (\AA) \\
\cline{1-6}
 nodal & $2.5$ &  $< 80$ & $> 1000$ & $\rightarrow 0$ & $\rightarrow
 \infty$ \\
 antinodal & $0.5$ & 80 & 250 & 4.3 & 165 \\

\end{tabular}
\end{ruledtabular}
\footnotetext[1]{The experimental error is $\sim 20 - 30$\% . }
\footnotetext[2]{The coherence length is calculated from $\xi_0 =
v_F/(\sqrt{12}\,\Delta)$. }
\label{tab:parms}
\end{table}

%
% Problems with the mean free-path...
%
In terms of the mean free path $l$ and the superconducting coherence length
$\xi_0$, the clean and dirty limits are expressed as $l \gg \xi_0$ and $l \leq
\xi_0$, respectively. Allowing that the mean free path is simply the Fermi
velocity times the scattering time, $l = v_F \tau$, and that the coherence
length for an isotropic gap in the weak-coupling regime is $\xi_0 \simeq v_F
/(\sqrt{12}\,\Delta)$ (Ref.~\onlinecite{benfatto02}), then the statement that
$1/\tau \approx 2\Delta_0$ and $l \approx \xi_0$ are roughly equivalent. In a
previous study of Nd$_{1.85}$Ce$_{0.15}$CuO$_4$, it was determined that $1/\tau
\approx 100$~cm$^{-1}$ just above $T_c$ (Ref.~\onlinecite{homes97}). Using an
average value for the Fermi velocity of $v_F = 2.2 \times 10^7$~cm/s
(Ref.~\onlinecite{allen88}) yields $l \approx 730$~\AA .  Given that $\xi_0
\simeq 80 - 90$~\AA\ in the electron-doped materials,\cite{anlage94,wilson06}
this appeared to justify the statement that this material was in the clean
limit, contradicting the present result.
%
% Anisotropic Fermi velocity
%
However, in addition to $1/\tau$, the anisotropy of the Fermi velocity is also
well documented in the high-temperature superconductors, and is found to vary
from $v_F \approx 2.5 - 2.7 \times 10^7$~cm/s along the nodal
direction,\cite{mesot99,zhou03} to $v_F \approx 0.5 \times 10^7$~cm/s along the
antinodal direction;\cite{damascelli06} while these values appear to be
remarkably universal, some sample and doping dependence is expected. Thus, the
mean free path determined from $1/\tau$ should be based on the antinodal $v_F$;
this yields a significantly smaller value of $l \simeq 250$~\AA .  Based on the
anisotropy of the Fermi velocity alone, the mean free path along the nodal
direction will be considerably larger.
The estimated value $\Delta_0$ in the present work is $\simeq 4.3$~meV; using
the value for the antinodal Fermi velocity yields $\xi_{0,calc} \simeq 165$~\AA
, which is about twice as large as the commonly quoted experimental values of
$\xi_0 = 80 - 90$~\AA ; the lack of perfect agreement may be partially due to
the uncertainty in $v_F$, but it is more likely a result of the na\"{\i}ve
approach taken to calculate $\xi_0$. In the nodal direction, the gap vanishes
and the coherence length diverges. These results are summarized in
Table~\ref{tab:parms}.

What these calculations indicate is that the assertion that $l \gg \xi_0$
arises only if the mean free path along the nodal direction is compared with
the coherence length in the antinodal direction; if the nodal and antinodal
directions are considered as separate cases, then it is indeed the case that $l
\approx \xi_0$. Thus, the result that $1/\tau \approx 2\Delta_0$ is in fact
consistent with the statement that $l \approx \xi_0$, where it is understood
that we are referring to the antinodal direction.  In fact, the statement
$1/\tau \approx 2\Delta_0$ should be considered more robust because it does not
rely on $v_F$. This implies that the material is not in the clean limit. Note
that this statement should be qualitatively correct along the nodal direction
as well, but because precise values of $l$ and $\xi_0$ [or $1/\tau$ and
$\Delta(\phi)$] are difficult to determine, this statement is somewhat
speculative.

%
% Conclusions
%
\section{Conclusions}
The {\it ab}-plane optical properties of single crystal
Pr$_{1.85}$Ce$_{0.15}$CuO$_4$ ($T_c \simeq 20$~K) have been examined above and
below $T_c$.  In the normal state just above $T_c$, the coherent part of the
optical conductivity may be described by a simple Drude component with
$\omega_{pd} \simeq 13\,000$~cm$^{-1}$ and $1/\tau \simeq 80$~cm$^{-1}$.  It is
noted that the condition $\hbar/\tau \approx 2k_B T$ near $T_c$ observed in
this material is generally true for many other cuprate superconductors.  Below
$T_c$, the superconducting plasma frequency is estimated to be $\omega_{ps}
\simeq 7800$~cm$^{-1}$, yielding a penetration depth of $\lambda \simeq
2000$~\AA ; when combined with the optical estimate for the dc resistivity
$\sigma_{dc}$ just above $T_c$, this material falls on the scaling relation
$\rho_s \propto \sigma_{dc} T_c$ recently proposed for the cuprate
superconductors.\cite{homes04,homes05}  The estimate for the superconducting
gap maximum $2\Delta_0 \simeq 70$~cm$^{-1}$ is in good agreement with previous
results,\cite{blumberg02,zimmers04} and is consistent with the view that the
superconducting energy gap is most likely nonmonotonic {\it d} wave.  The
result that $1/\tau \approx 2\Delta_0$ implies that this material is not in the
clean limit, a self-consistent result that by its own nature allows for the
direct observation of the superconducting energy gap.

%
% Acknowledgements
%
%\begin{acknowledgments}
We would like to acknowledge helpful discussions with D.~N.~Basov,
A.~V.~Chubukov, A.~J.~Millis, T.~Valla and N.~L.~Wang.
This work was supported by the Office of Science, U.S. Dept. of Energy, under
contract number DE-AC02-98CH10886; work in Maryland is supported by the NSF
contract DMR-0352735.
%\end{acknowledgments}

%
%%%%%%%%%%%%%%%%%%%%%%%%%%%%%%%%%%%%%%%%%%%%%%%%%%%%%%%%%%%%%%%%%%%%%%%%%%%%%%
%
% References
%
\bibliography{pcco}

\end{document}